\documentclass[aps,prd,onecolumn,nofootinbib,superscriptaddress]{revtex4}
\usepackage{graphicx}
\pdfoutput=1
\usepackage{epsfig}
\usepackage{amsmath}
\usepackage{amsfonts}
\usepackage{amssymb}
\usepackage{color}
\usepackage{dcolumn}
\usepackage{hyperref}
\usepackage{multirow}
\usepackage{booktabs}
\usepackage{MnSymbol,wasysym}
\usepackage{braket,diagbox}
\usepackage{eurosym}
\usepackage{calrsfs}
\providecommand{\U}[1]{\protect\rule{.1in}{.1in}}

\newcommand{\f}{\begin{equation}}
\newcommand{\ff}{\end{equation}}
\newcommand{\fa}{\begin{eqnarray}}
\newcommand{\ffa}{\end{eqnarray}}

\begin{document}
\baselineskip=0.4 cm
\title{ Strong gravitational lensing effects around rotating regular black holes}

\author{Ming-Yu Guo}
\affiliation{School of Mathematics, Physics and Statistics, Shanghai University of Engineering Science, Shanghai 201620, China}
\affiliation{Center of Application and Research of Computational Physics, Shanghai University of Engineering Science, Shanghai 201620, China}

\author{Meng-He Wu}
\email{mhwu@njtc.edu.cn} 
\affiliation{College of Physics and Electronic Information Engineering, Neijiang Normal University, Neijiang 641112, China}

\author{Hong Guo}
\affiliation{Escola de Engenharia de Lorena, Universidade de S\~ao Paulo, 12602-810, Lorena, SP, Brazil}

\author{Xiao-Mei Kuang}
\email{xmeikuang@yzu.edu.cn}
\affiliation{ Center for Gravitation and Cosmology, College of Physical Science and Technology, Yangzhou University, Yangzhou 225002, China}

\author{Fu-Yao Liu}
\affiliation{School of Mathematics, Physics and Statistics, Shanghai University of Engineering Science, Shanghai 201620, China}
\affiliation{Center of Application and Research of Computational Physics, Shanghai University of Engineering Science, Shanghai 201620, China}

\begin{abstract}
In this letter, we investigate the strong gravitational lensing effects around two classes of rotating regular black holes, which behave as non-singular Minkowski core at the center. Starting from the null geodesic in the equatorial plane of the regular black holes, we analyze the deflection angle for both prograde photons and retrograde photons, which are found to significantly shift from that for Kerr black hole. Then we suppose the rotating regular black holes as the supermassive M87* and SgrA* black holes, respectively, and evaluate the lensing observables such as the image position, separation, magniﬁcation and the time delays between the relativistic images for the black holes. In both cases, the time delay differences between the regular black holes and Kerr black hole seem to be measurable, especially for the M87* black hole, but the deviations of outermost relativistic image and the asymptotic relativistic image for the two rotating regular black holes from those for Kerr black hole are less than 10 $\mu as$. This means that it is difficult to probe the regularity of the black hole via the current observations, but we can expect the next generation Event Horizon Telescope to test more precise properties of strong field regime.

\end{abstract}

\maketitle
\tableofcontents

\section{Introduction}
{The latest observational development on gravitational waves \cite{LIGOScientific:2016aoc,LIGOScientific:2018mvr,LIGOScientific:2020aai} and supermassive black hole shadows \cite{EventHorizonTelescope:2019dse,EventHorizonTelescope:2019ggy,EventHorizonTelescope:2019ths,
EventHorizonTelescope:2022xnr,EventHorizonTelescope:2022xqj} further demonstrate the great success of Einstein's general relativity (GR). Meanwhile, the uncertainties in the data also leave some
space for alternative theories of gravity, which further triggers physicists' interest to probe various theoretical predictions.
Moreover, the challenges faced by GR, including the explanation of the universe expansion history, the large scale structure and the understanding of the quantum gravity, remind us that a more general theory of gravity is required. Thus, various generalized theories of gravity have been proposed, which essentially provide us richer frameworks to further understand the nature of gravity. In particular, 
the regular (non-singular) black holes, which have no essential singularity, are proposed to resolve the singularity problem in GR. In our knowledge, 
there exists two independent ways to construct regular black holes. One is to directly solve equations of motions in extended theories containing special sources, see for examples \cite{dymnikova1992vacuum, Nicolini:2005vd,Balakin:2016mnn,Roupas:2022gee} and references therein, in which the regular black hole behaves semiclassically. The other is to obtain regular black hole as quantum corrections to the classical black hole with singularity, see for examples \cite{Borde:1996df, Bonanno:2000ep,Gambini:2008dy,Perez:2017cmj,Brahma:2020eos}. In this case, the regular black holes usually exhibit quantum behaviors, thus it is believed that the black hole singularity could be removed or avoided by considering the quantum effects of gravity. So the regular black hole can be one of powerful tools to study the classical
limit of quantum black holes before a well-defined quantum gravity theory is established. Readers can refer to \cite{Torres:2022twv,Lan:2023cvz} for reviewing the development of regular black hole.

Gravitational lensing is a consequence of GR, and it plays a foundational and influential role in theoretical physics, astrophysics and astronomy. Firstly,
since the image of the supermassive M87* black hole was published, the shadow of black holes has received considerable attention. The photon region formed because of strong gravitational lensing is the core part in the black hole shadow. The existence of unstable photon regions gives the possibility for us to observe the black hole directly, and the photons that escape from the spherical orbits form the boundary of the dark silhouette of the black hole \cite{Cunha:2018acu,Perlick:2021aok,Chen:2022scf}. Secondly, the gravitational lensing can be a powerful tool to probe distant celestial objects (galaxies, quasars and stars, etc.) and possible entities such as dark matter, dark energy, and gravitational waves, and it could be helpful for us to comprehensively understand various astronomical phenomena. 
Last but not least, the gravitational lensing in the strong gravity regime of black hole is an effective way to study the near horizon properties of black hole. The lens equation for the strong field limit of Schwarzschild black hole was firstly introduced numerically in \cite{Virbhadra:1999nm,Virbhadra:2002ju}. Then, in \cite{Bozza:2002zj}, Bozza proposed an analytical logarithmic expansion method for strong field lensing, which was lately extended into a general asymptotically flat spacetime \cite{Tsukamoto:2016jzh}. The analytical method was then employed to connect various lensing observables for static and spherically symmetric
spacetimes \cite{Bozza:2002zj,Perlick:2003vg}, which was then extensively studied in \cite{Bozza:2003cp,Beckwith:2004ae,Eiroa:2003jf,Whisker:2004gq,Sarkar:2006ry,Chen:2009eu,Wei:2011nj,Panpanich:2019mll,Zhao:2016kft,Lu:2021htd,Shaikh:2018oul,Babar:2021nst,Kumar:2020sag,Tsukamoto:2021caq,Javed:2019qyg,Kuang:2022xjp,Kuang:2022ojj,Kumar:2019pjp,Kumar:2021cyl,Pietroni:2022cur,Liu:2023xtb,Xie:2024dpi,QiQi:2023nex,Feleppa:2024vdk,Zhao:2024elr}.
The technical development from Event Horizon Telescope make it possible to directly observe and explore the strong gravity regime, so we could understand the properties of black holes from the strong gravitational lensing effect. 

Especially, the lensing observables may serve as diagnosis to disclose some fundamental problem on the nature of gravity, such as alternative theories of gravity, the candidates for dark matter and dark energy and the quantum effect of gravity.
In the multi-message era in astrophysics, the studies on theoretical predictions from lensing observables, together with gravitational wave, shadow and image, quasi-normal mode and particle's motion around black holes etc., could further testify the existence of regular black hole.

The aim of this letter is to study the strong gravitational lensing effects around a rotating regular black hole with Minkowski core, which is the rotating counterpart of regular black hole with spherical symmetry proposed in \cite{Ling:2021olm}. }
We shall focus on the cases with $(\gamma= 2/3, n = 2)$ and $(\gamma= 1, n = 3)$ which respectively correspond to Bardeen black hole \cite{bardeen1968proceedings} and Hayward black hole \cite{Hayward:2005gi} with Minkowski core as the spinning parameter vanishes \cite{Ling:2021olm}.
Based on their deflection angle of strong gravitational lensing, we will presuppose the rotating regular black holes as supermassive M87* and SgrA* black holes and evaluate the lensing observables, such as the image position, separation, magnification and the time delays between the relativistic images. We will mainly analyze the influence of the parameters of the two rotating regular black holes, and distinguish the differences of various lensing observables between the two cases and their shift from Kerr black hole. It is noted that though current observational techniques for detecting deviations from the Kerr black hole model have advanced, they still face limitations because the deviations should depend significantly on detecting subtle effects in the vicinity of black hole, such as the size and shape of the shadow, lensing observable, the orbits of nearby stars, and gravitational wave and so on. Current observations and data analysis precessing on these phenomena still have challenges in resolution, sensitivity, and reliance on model assumptions. Nevertheless, it is optimistic because future advancements in observational technology, especially those of the EHT and the integration of multi-messenger data, could significantly reduce these limitations, enabling more precise tests for deviations from the Kerr black hole model. So we hope that our preliminary estimation on the derivation from theoretical aspect could give some insight for future test.

The remaining of this letter is organized as follows. In section \ref{sec:background}, we will review the rotating Kerr-like regular black hole with Minkowski core. In section \ref{sec:angle}, we shall study the light deflection angles lensed by the aforementioned two types of rotating regular black hole in the strong field regime. We then suppose the rotating regular black holes as the supermassive M87* and SgrA* black holes and evaluate various lensing observables in section \ref{sec:observables}. The last section is our conclusion and discussion.

\section{Regular black holes with Minkowski core}\label{sec:background}

In this section, we briefly review the regular black holes proposed in \cite{Ling:2021olm} , whose spherically symmetric metric is as follows
\begin{equation}
   {ds}^2=-f\left(r\right){dt}^2+f({r})^{-1}{dr}^2+r^2 \left({d\theta }^2+ \sin ^2(\theta )\right){d\phi }^2,\label{eq:3}
\end{equation}
where $f(r)=1-\frac{2 m(r)}{r}$ and $m(r)$ takes the form as
\begin{equation}
m(r)=M e^{- g^n M^\gamma/{r^n}}. \label{eq:4}
\end{equation}
When the parameter $g=0$, it is transformed into a static spherically symmetric Schwarzschild black hole. The presence of the parameter $g$ shifts the position of the maximum Kretschmann scalar curvature to a larger radius, which could have significant implications for the spacetime geometry and the behavior of gravitational properties. Moreover, the parameter $g$ also impacts the Hawking temperature, initially increasing it before subsequently decreasing it. This fluctuation in temperature may indicate alterations in the thermodynamic characteristics of the black hole, including its entropy and heat capacity, and could have implications for the stability and evaporation dynamics of the black hole \cite{Ling:2021olm}.

Those regular black holes have the following significant characteristics. Firstly, the exponentially diminishing Newton potential results in a non-singular Minkowski core at the center of the black hole, a concept initially proposed in \cite{xiang2013singularities}. Secondly, within the constraint $3/n<\gamma<n$ and $n\geqslant2$, the Kretschmann scalar curvature remains sub-Planckian regardless of the mass of black hole. Additionally, the correspondence between those regular black holes and those featuring asymptotically de Sitter cores has been highlighted in \cite{Ling:2021olm}, where the mass function $m(r)$ is defined as follows
\begin{equation}
m(r)=-\frac{M r^{\frac{n}{\gamma}-1}}{\left(r^n+ g \gamma M^\gamma\right)^{1 / \gamma}}.
\end{equation}

\begin{figure}[ht]
    \centering
    \includegraphics[width=0.3\linewidth]{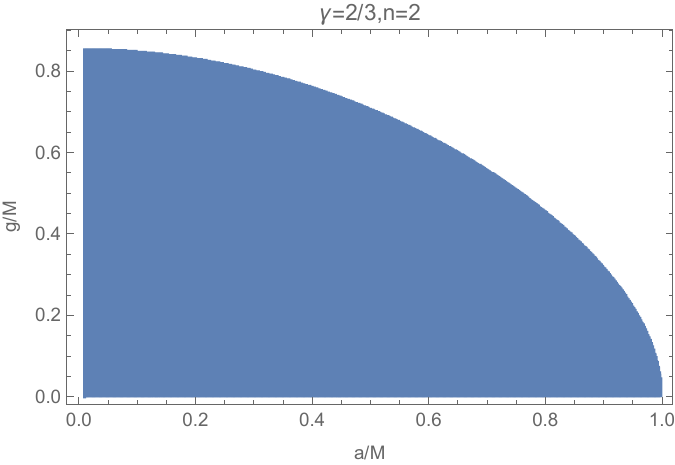}\hspace{1cm}
     \includegraphics[width=0.3\linewidth]{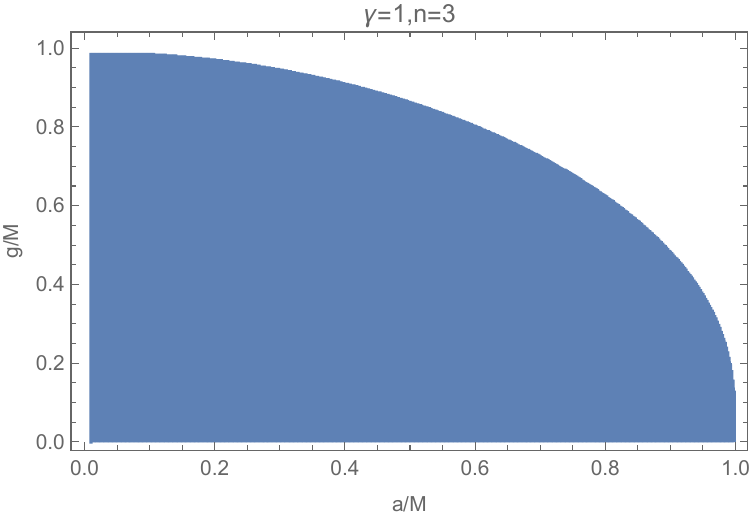}
    \caption{The figure shows that different spins $a$ have different ranges $g$ which  decreases as $a$ increases for the regular black hole with $\gamma= 2/3, n = 2$ (left panel) and $\gamma = 1, n = 3 $ (right panel), respectively.}
    \label{figa3}
\end{figure}

Specially, when $\gamma= 2/3, n = 2$, it results in the Bardeen black hole, whereas when $\gamma = 1, n = 3 $, it leads to Hayward black hole. In this letter, we concentrate on the black hole described by Eq. \ref{eq:4}, specifically for $\gamma = 2/3$ and $n = 2$, and for $\gamma = 1$ and $n = 3$, to ensure that the maximum value of the Kretschmann scalar curvature is independent of the black hole mass, as discussed in \cite{Ling:2021olm}.
Given that astrophysical black holes in our universe typically exhibit rotation, the authors of \cite{newman1965note} derived the rotating equivalent of the static black hole. This rotating black hole is both stationary and axisymmetric, and it is expressed in Boyer-Lindquist coordinates as follows
\begin{equation}\label{eq:16}
\begin{aligned}
d s^2 & =-\left(1-\frac{2 m(r) r}{\Sigma}\right) d t^2-\frac{4 a m(r) r \sin ^2 \theta}{\Sigma} d t d \phi+\frac{\Sigma}{\Delta} d r^2 \\
& +\Sigma d \theta^2+\left(r^2+a^2+\frac{2 a^2 m(r) r \sin ^2 \theta}{\Sigma}\right) \sin ^2 \theta d \phi^2,
\end{aligned}
\end{equation}
with $\Sigma = r^2 + a^2 \cos^2 \theta$ and $\Delta (r)=a^2-2 m( r) r+r^2$, where $m(r)$ takes the form of Eq.\ref{eq:4}.
The radius of these black hole horizons can be obtained by solving for the real positive root of $\Delta (r)=0$. To demonstrate the parameter region of the regular black hole, we show the plots in parameters space $(g,a)$ for $\gamma= 2/3, n = 2$ and $\gamma = 1, n = 3 $ in Fig.\ref{figa3}, where the parameters in the shadow regimes admit the existence of the black hole.

\begin{figure}[ht]
    \centering
    \includegraphics[width=0.25\linewidth]{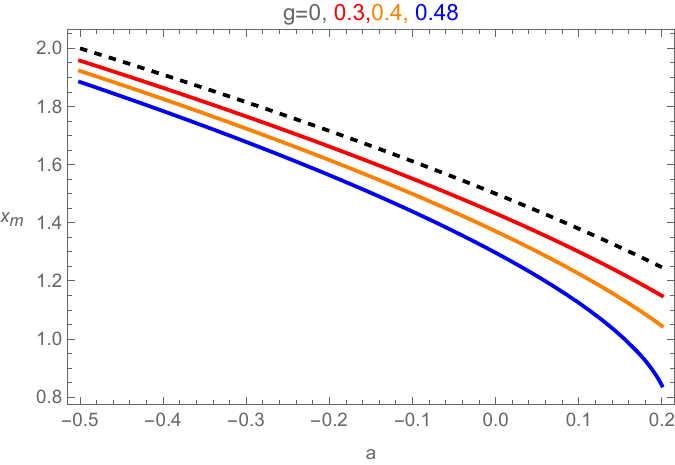}\hspace{1cm}
     \includegraphics[width=0.25\linewidth]{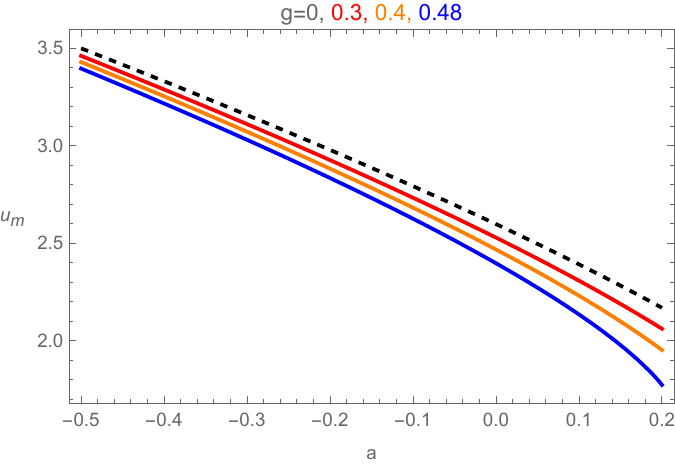}\\
       \includegraphics[width=0.25\linewidth]{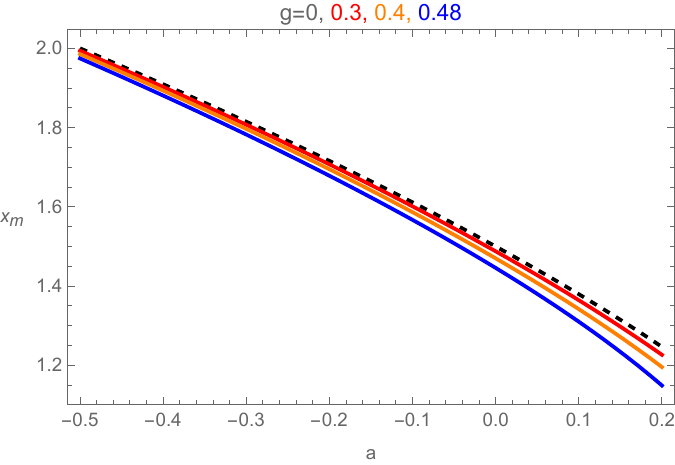}\hspace{1cm}
     \includegraphics[width=0.25\linewidth]{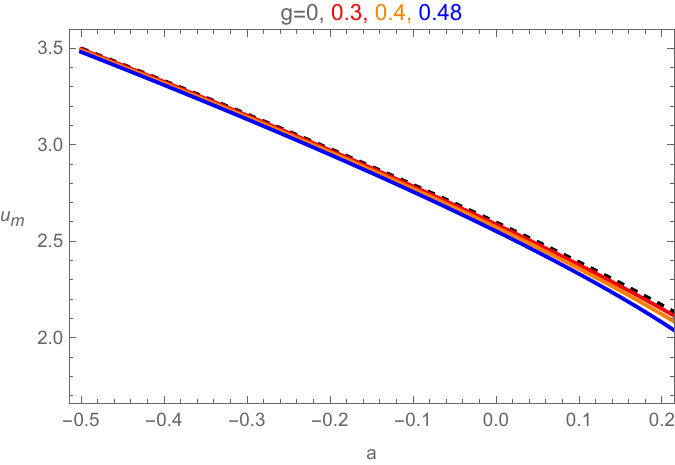}
    \caption{The radius of the photon circle $x_m$ and the critical impact parameter $u_m$ for the regular black hole with $\gamma= 2/3, n = 2$ (upper panel) and $\gamma = 1, n = 3 $ (lower panel). The dashed black curves describe the Kerr case.}
    \label{figa12}
\end{figure}

\section{Strong gravitational lensing effect}\label{sec:angle}

In this section, our research will focus on the gravitational lensing phenomena of the rotating regular black holes, and thus investigate the effect of the parameter $g$ on the lensing observations.

\subsection{Deflection angle of strong gravitational lensing }

\subsubsection{Null geodesic in the equatorial plane}
To investigate the gravitational lensing effect, we will adhere to the methodology outlined in \cite{Bozza:2002zj}, focusing on light rays in the equatorial plane. We rescale all quantities $r$, $a$, and $t$ in units of $2M$ with $M = 1/2$, and substitute $x$ for $r$. The metric Eq.\ref{eq:16} projected onto the equatorial plane can then be represented as
\begin{equation}
d s^2=-A(x) d t^2+B(x) d x^2+C(x) d \varphi^2-D(x) d t d \varphi,\label{eq:46}
\end{equation}
where
\begin{equation}
\begin{array}{ccc}
A(x)=1-\frac{2 m(x)}{x},&B(x)=\frac{x^2}{a^2-2 x m(x)+x^2},&\\
C(x)=\frac{2 a^2 m(x)}{x}+a^2+x^2,&D(x)=\frac{4 a m(x)}{x} ,&\label{eq:47}
\end{array}
\end{equation}
with $m(x)$ taking the form of Eq.\ref{eq:4}.
The Lagrangian system for a photon is $\mathcal{L}=\frac{1}{2}(\frac{ds}{d\tau})^2=\frac{1}{2} g_{\mu \nu } \dot{x}^{\mu } \dot{x}^{\nu }$ and Hamilton-Jacobi equation is 
\begin{equation}
\mathcal{H}=-\frac{\partial S}{\partial \tau }=\frac{1}{2}g^{\alpha \beta }\frac{\partial S}{\partial x^{\alpha }}\frac{\partial S}{\partial x^{\beta }},\label{eq:18}
\end{equation}
where $\mathcal{H}$ is the canonical Hamiltonian, $S$ is the Jacobi action, and $\tau$ is an affine parameter along the geodesics. Because the metric Eq.\ref{eq:16} is a Kerr-like black hole form, we have the Killing vector fields $\partial_t$ and $\partial_\phi$, and there are two conserved quantities
\begin{equation}
E=-p_t=- g_{tt}\dot{t}- g_{t\phi }\dot{\phi }, \quad L_z=-p_{\phi }=g_{\phi \phi }\dot{\phi } + g_{\phi t}\dot{t}.\label{eq:19}
\end{equation}

According to Hamiltonian equation Eq.\ref{eq:18}, two conserved quantities Eq.\ref{eq:19} that determine the trajectory of the photon can be obtained. And by using the Hamilton-Jacobi method \cite{carter1968global} to analyze the trajectory of the photon, and denoting $L_z/E$, we derive the zero geodesic equation as
\begin{equation}
\begin{aligned}
\dot{t} & =\frac{4 C-2 u D}{4 A C+D^2} ,\\
\dot{\phi} & =\frac{2 D+4 A u}{4 A C+D^2} ,\\
\dot{x} & = \pm 2 \sqrt{\frac{C-D u-A u^2}{B\left(4 A C+D^2\right)}},\label{eq:48}
\end{aligned}
\end{equation}
where the dot indicates the derivative with respect to the affine parameter and the $+$ and $-$ denote that the photon travels in the counterclockwise and clockwise direction, respectively. We can establish the effective potential governing radial motion as follows
\begin{equation}
V_{\mathrm{eff}}=\frac{4\left(C-D u-A u^2\right)}{B\left(4 A C+D^2\right)},\label{eq:49}
\end{equation}
which determine various photon orbits. When $V_{\mathrm{eff}}(x=x_0) = 0$, the light ray emitted from the light source may deflect towards the black hole at the radius $x_0$, which is indicated as the minimum approach distance to the black hole, and then proceed towards the observer.
The impact parameter can be obtained
\begin{equation}
L=u\left(x_0\right)=\frac{-D\left(x_0\right)+\sqrt{4 A\left(x_0\right) C\left(x_0\right)+D\left(x_0\right)}}{2 A\left(x_0\right)}.\label{eq:50}
\end{equation}
When light approaches the photon circle of radius $x_0=x_m$, the deflection angle diverges, as determined by the following
\begin{equation}
V_{\mathrm{eff}}=\left.\frac{d V_{\mathrm{eff}}}{d x}\right|_{\left(x_0=x_m\right)}=0.\label{eq:51}
\end{equation}
In addition, unstable photon circles must further satisfy the following condition
\begin{equation}
\left.\quad \frac{d^2 V_{\mathrm{eff}}}{d x^2}\right|_{\left(x_0=x_m\right)}<0.  \label{eq:52}  
\end{equation}
Hence, the photon orbit radius $x_m$ is determined as the largest root of the following equation
\begin{equation}
A \frac{d C}{d x}- \frac{d A}{d x} C+L\left(\frac{d A}{d x} D-A \frac{d D}{d x}\right)=0.\label{eq:53}
\end{equation}
The critical impact parameter $u_m$ can be computed by $u$ in the case $x = x_m$, namely $u_m \equiv u\left(x_m\right)$.
In Fig.\ref{figa12}, We plot $x_m$ and $u_m$ as functions of $a$ in the case of regular black holes with $\gamma= 2/3, n = 2$ and $\gamma = 1, n = 3 $ for various parameters $g$.
{We observe that in both cases, the radius of the photon circle $x_m$ and critical impact parameter $u_m$ decrease as the spin $a$ increases, which is similar to that in Kerr black hole \cite{islam2021strong}. However, comparing with Kerr case $g=0$ represented by the black dotted curve, it is obvious that for the regular black hole, both $x_m$ and $u_m$ decrease with the increasing of parameter $g$. This phenomena will be reflected in the effect of $g$ on the light deflection angle and the lensing observables as we will study soon.}
It is also observed that $x_m$ and $u_m$ for the regular black hole with $\gamma= 2/3, n = 2$ changes more significantly than those with $\gamma= 1, n = 2$.

\subsubsection{Deflection angle}
\begin{figure}[ht]
    \centering
    \includegraphics[width=0.25\linewidth]{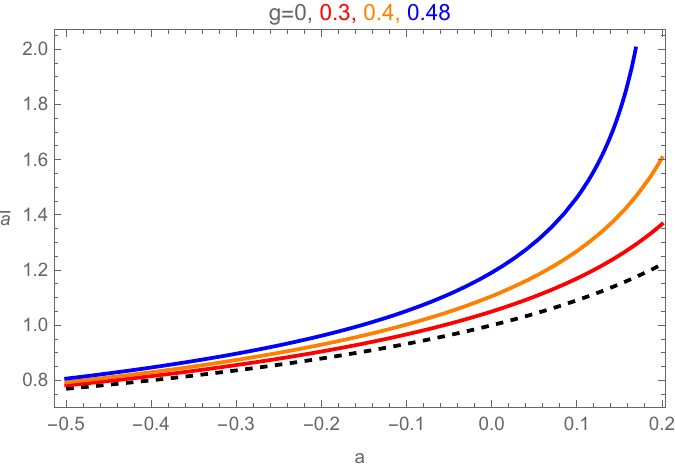}\hspace{1cm}
     \includegraphics[width=0.25\linewidth]{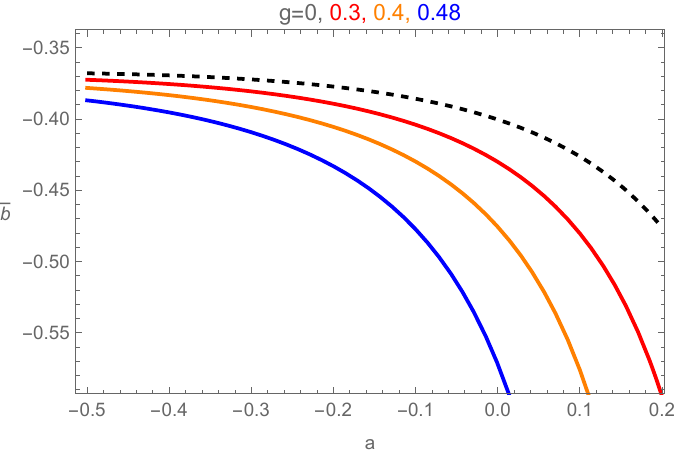}\\
       \includegraphics[width=0.25\linewidth]{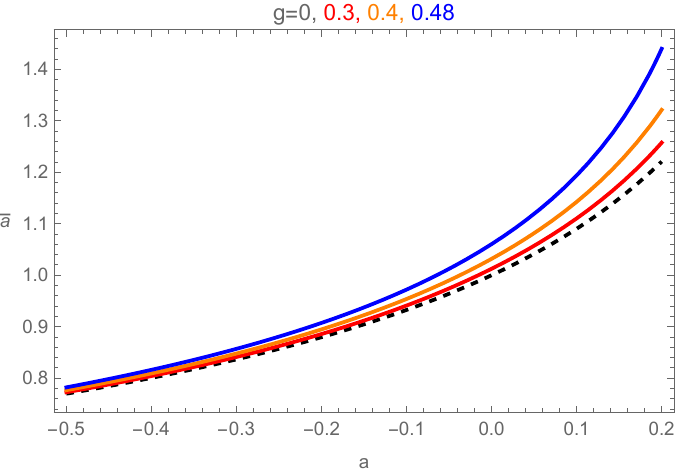}\hspace{1cm}
     \includegraphics[width=0.25\linewidth]{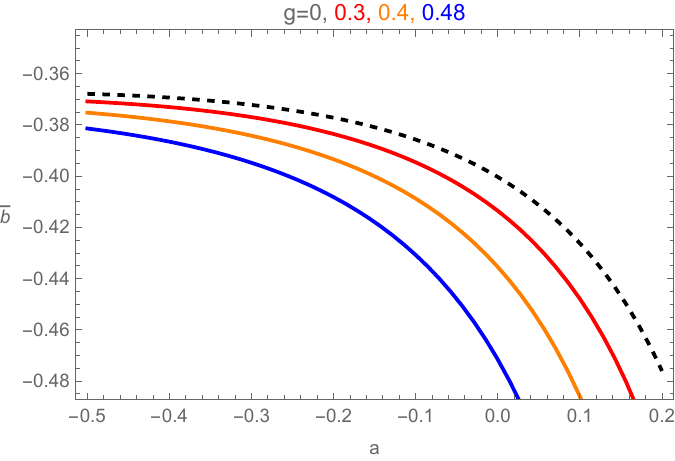}
    \caption{The behaviors of the lensing coefficients $\bar a$ and $\bar b$ for the regular black hole with $\gamma= 2/3, n = 2$ (upper panel) and $\gamma = 1, n = 3 $ (lower panel). The dashed black curves describe the Kerr case.}
    \label{figa13}
\end{figure}

According to \cite{carter1968global}, the deflection angle of light by the rotating black hole at $x_0$ is determined as:
\begin{equation}
\alpha_D(x_0)=I(x_0)-\pi,  \label{eq:54}
\end{equation}
where
\begin{equation}
I\left(x_0\right)=2 \int_{x_0}^{\infty} \frac{d \varphi}{d x} d x=2 \int_{x_0}^{\infty} \frac{\sqrt{A_0 B}(2 A L+D)}{\sqrt{4 A C+D^2} \sqrt{A_0 C-A C_0+L\left(A D_0-A_0 D\right)}} d x .\label{eq:55}
\end{equation}
However, the integral usually cannot be solved out in an explicit form. Adopting the methods in \cite{bozza2001strong,Tsukamoto:2016jzh,Bozza:2002zj}, we can expand the deflection angle near the strong deflection limit of the photon circle to simplify this complex integral, and provide the analytical formula for the deflection angle. To proceed, we introduce the auxiliary variable $z = \frac{A-A_0}{1-A_0}$, the integral is then rewritten as
\begin{equation}
I\left(x_0\right)=\int_0^1 R\left(z, x_0\right) f\left(z, x_0\right) d z,  \label{eq:56}
\end{equation}
where
\begin{equation}
R\left(z, x_0\right)=\frac{2\left(1-A_0\right)}{A^{\prime}} \frac{\sqrt{B}\left(2 A_0 A L+A_0 D\right)}{\sqrt{C A_0} \sqrt{4 A C+D^2}},
\label{eq:57}
\end{equation}
\begin{equation}
f\left(z, x_0\right)=\frac{1}{\sqrt{A_0-A \frac{C_0}{C}+\frac{L}{C}\left(A D_0-A_0 D\right)}} .
\label{eq:58}
\end{equation}
The function $R\left(z, x_0\right)$ remains well-behaved for any values of $z$ and $x_0$, while the function $f\left(z, x_0\right)$ diverges as $z \rightarrow 0$. To address this divergence, we expand the expression for the square root of $f\left(z, x_0\right)$, yielding
\begin{equation}
f\left(z, x_0\right) \sim f_0\left(z, x_0\right)=\frac{1}{\sqrt{\mathfrak{m}\left(x_0\right) z+\mathfrak{n}\left(x_0\right) z^2}},\label{eq:59}
\end{equation}
where $\mathfrak{m}\left(x_0\right)$ and $\mathfrak{n}\left(x_0\right)$ are the coefficients of Taylor expansion.
Then we can obtain the strong-field limit of the deflection angle \cite{Bozza:2002zj,Tsukamoto:2016jzh} as
\begin{equation} \label{eq:60}
\alpha_D(u)=-\bar{a} \log \left(\frac{L}{u_m}-1\right)+\bar{b}+\mathcal{O}\left(u-u_m\right),
\end{equation}
with the strong deflection coefficients $\bar{a}$ and $\bar{b}$ are
\begin{equation}
\bar{a}=\frac{R\left(0, x_m\right)}{2 \sqrt{\mathfrak{n}_m}}, \quad \text { and } \quad \bar{b}=-\pi+b_D+b_R+\bar{a} \log \frac{\bar{c} x_m^2}{u_m}, \label{eq:61}
\end{equation}
where
\begin{equation}
b_D=2 \bar{a} \log \frac{2\left(1-A_m\right)}{A_m^{\prime} x_m}, \quad b_R=\int_0^1\left[R\left(z, x_m\right) f\left(z, x_m\right)-R\left(0, x_m\right) f_0\left(z, x_m\right)\right] d z \label{eq:62}
\end{equation}
and $\bar c$ is defined by the coefficient in Taylor expansion
\begin{equation}
u-u_m=\bar c(x_0-x_m)^2. \label{eq:63}
\end{equation}
As depicted in Fig.\ref{figa13}, we illustrate the strong deflection coefficients $\bar{a}$ and $\bar{b}$ as functions of the spin $a$ for selected values of $g$ in the case of regular black holes with $\gamma= 2/3, n = 2$ and $\gamma = 1, n = 3 $, respectively. It can be observed that $\bar{a}$ increases as the spin $a$ increases, and with the parameter $g$ increasing, $\bar{a}$ consistently grows up. In detail, the effect of parameter $g$ on prograde photons ($a>0$) becomes more pronounced with increasing spin. Conversely, for retrograde photons ($a<0$), the lensing coefficients $\bar{a}$ closely resemble those of the Kerr black hole. However, as shown in the right part of Fig.\ref{figa13}, the spin $a$ has a diminishing effect on $\bar{b}$, and an increase in the parameter $g$ results in a smaller $\bar{b}$.

\begin{figure}[ht]
    \centering
    \includegraphics[width=0.25\linewidth]{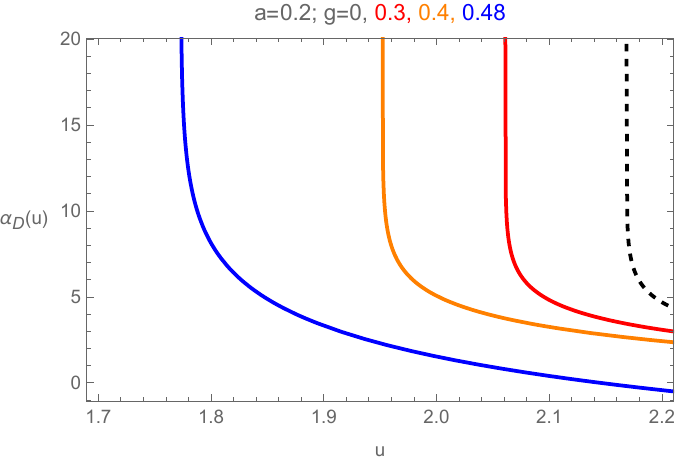}\hspace{1cm}
     \includegraphics[width=0.25\linewidth]{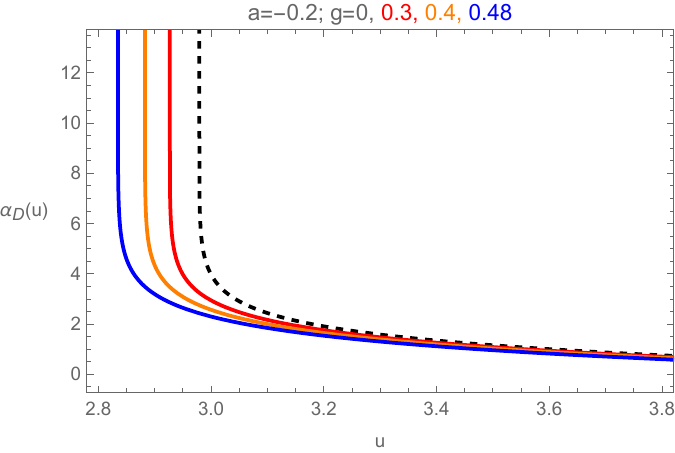}\\
         \includegraphics[width=0.25\linewidth]{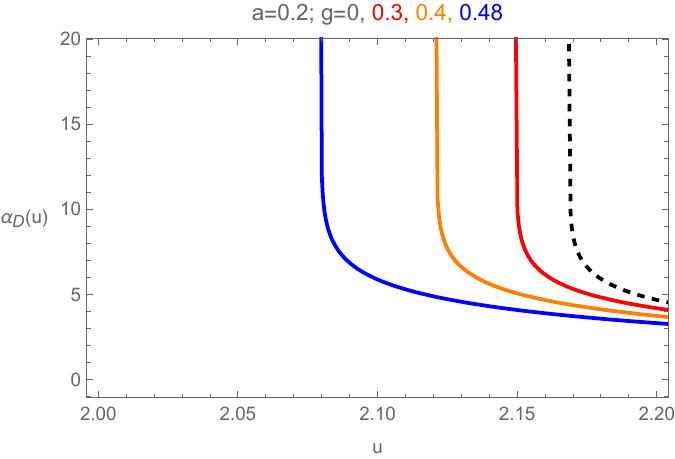}\hspace{1cm}
     \includegraphics[width=0.25\linewidth]{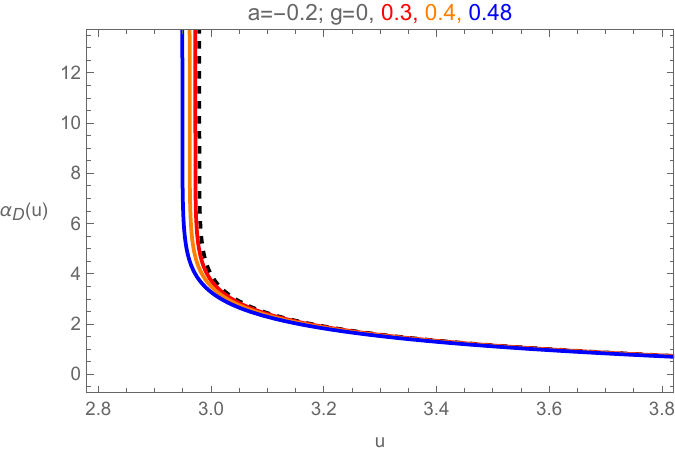}
    \caption{The deflection angle $\alpha _D$ is plotted as a function of $u$ with various values of $g$ for the regular black hole with $\gamma= 2/3, n = 2$ (upper panel) and $\gamma = 1, n = 3 $ (lower panel).
}
    \label{figa14}
\end{figure}

Furthermore, in Fig.\ref{figa14}, we plot the relationship between the deflection angle $\alpha_D(u)$ and $u$. It is noticeable that the deflection angle decreases monotonically as $u$ increases as expected. {Nevertheless, recalling the condition of strong field limit, the strong lensing deflection angles are only effective when $u$ approaches the corresponding $u_m$ for the parameters shown in Fig.\ref{figa12}. Our results indicate that the light deflection angle for the rotating regular black holes will be smaller than that for Kerr black hole.} In addition, we also observe that the deflection angle $\alpha_D(u)$ of a regular black hole with $\gamma= 2/3, n = 2$ changes more significantly than that with $\gamma= 1, n = 2$.

\section{Lensing observables by supermassive black holes}\label{sec:observables}

Having the deflection angles of light, we shall study various strong lensing observables of the rotating regular black holes, and checking their deviation from those of Kerr black hole in GR. To this end, we first review the general formulas of the lensing observables. Then we assume the rotating regular black holes as the supermassive black hole to evaluate the observables. 

\subsection{Formulas of lensing observables}

\begin{figure}[ht]
    \centering
    \includegraphics[width=0.45\linewidth]{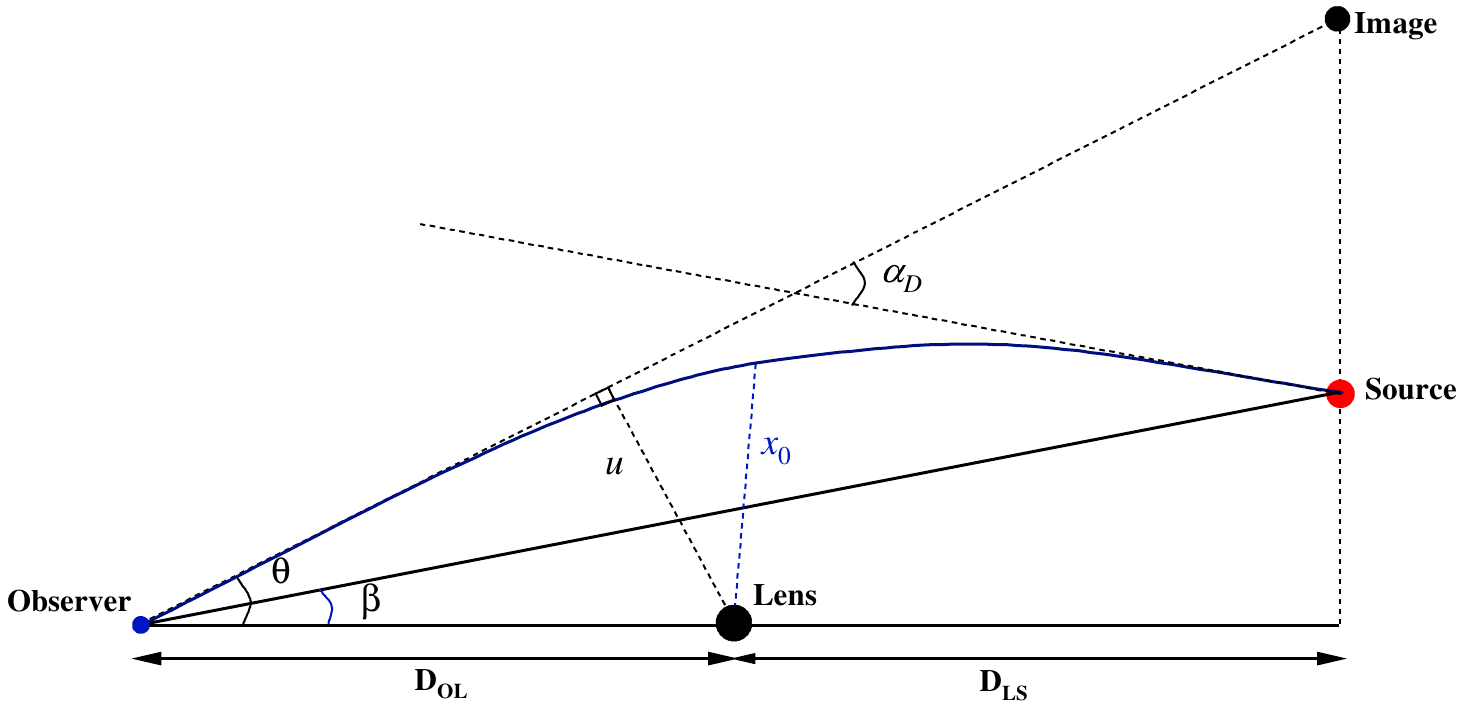}
    \caption{ The schematic diagram  of the gravitational lensing.}
    \label{figa15}
\end{figure}

To derive the equation for small deflection angles, we assume that both the light source and the observer are far from the black hole (lens). The angular distance between the light source and the black hole can be expressed as \cite{Bozza:2001xd}
\begin{equation}
\beta=\theta-\frac{D_{LS}}{D_{OS}}\Delta \alpha_n ,\label{eq:64}
\end{equation}
where $\theta $ is the angular distance between the image and the black hole, $D_{LS}$ is the distance between the lens and the light source, and $D_{OS}$ is the distance between the observer and the light source. $\Delta \alpha_n = \alpha_D-2n\pi$ represents the deviation of the deflection angle from $2n\pi$, indicating that the light source orbits the black hole multiple times before reaching the observer. The schematic diagram of the gravitational lensing is shown in Fig.\ref{figa15}.

According to Eqs.\ref{eq:60} and \ref{eq:64}, the position of the $n-th$ relativistic image is approximated as \cite{Bozza:2002zj}
\begin{equation}
\theta_n=\theta_n^0+\frac{u_m e_n\left(\beta-\theta_n^0\right) D_{O S}}{\bar{a} D_{L S} D_{O L}} ,\label{eq:65}
\end{equation}
with
\begin{equation}
e_n=\exp \left(\frac{\bar{b}-2 n \pi}{\bar{a}}\right) ,  \label{eq:66}
\end{equation}
where $\theta^0_n $ is the image position corresponding to $\alpha_D=2 n \pi$, $D_{OS}$ is the distance between the observer and the light source.

The brightness of the relativistic images will be amplified due to lensing. The magnification of the n-th image can be expressed as the ratio of the solid angle of the image to that of the light source \cite{Bozza:2002zj,islam2021strong}
\begin{equation}
\mu_n=\left.\left(\frac{\beta}{\theta} \frac{d \beta}{d \theta}\right)^{-1}\right|_{\theta_n^0}=\frac{u_m^2 e_n\left(1+e_n\right) D_{O S}}{\bar{a} \beta D_{L S} D_{O L}^2} . \label{eq:67}
\end{equation}
Due to the large value of $D_{O L}^2$, the actual brightness of the relativistic images is very faint, and it decreases as $n$ increases. Therefore, the first relativistic image is the brightest, and its brightness diminishes rapidly.
So the outermost image ($\theta_1$) is usually considered to be classified as one category of images, while all other images are categorized as $\theta_{\infty}$, representing the asymptotic position of a group of images in the limit $n \rightarrow \infty$. Combining the expressions for the deflection angle Eq.\ref{eq:60} and the lensing angle Eq.\ref{eq:64}, we can assess three observational quantities of the relativistic images, namely, the angular position of the asymptotic relativistic image ($\theta_{\infty}$), the angular separation between the outermost relativistic image and the asymptotic relativistic image ($s$), and the relative magnification of the outermost relativistic image compared to other relativistic images ($r_\mathrm{mag}$) \cite{Bozza:2002zj}
\begin{equation}
\begin{aligned}
& \theta_{\infty}=\frac{u_m}{D_{O L}}, \label{eq:68} \\
& s=\theta_1-\theta_{\infty}=\theta_{\infty} \exp \left(\frac{\bar{b}}{\bar{a}}-\frac{2 \pi}{\bar{a}}\right),  \\
& r_{\text {mag }}=\frac{\mu_1}{\sum_{n=2}^{\infty} \mu_n} \simeq \frac{5 \pi}{\bar{a} \log (10)}.
\end{aligned}
\end{equation}
{By calculating Eq.\ref{eq:68}, we can determine the distribution and size of the images for a known light source, and diagnosing the properties of the black hole lensing by connecting the observational results.}

\begin{figure}[ht]
    \centering
    \includegraphics[width=0.25\linewidth]{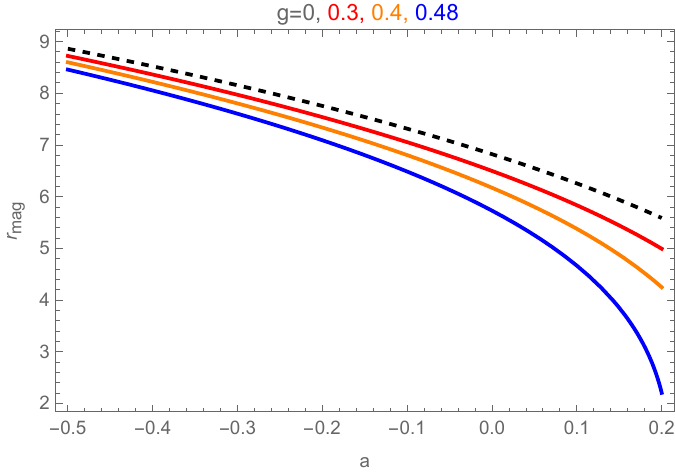}\hspace{1cm}
     \includegraphics[width=0.26\linewidth]{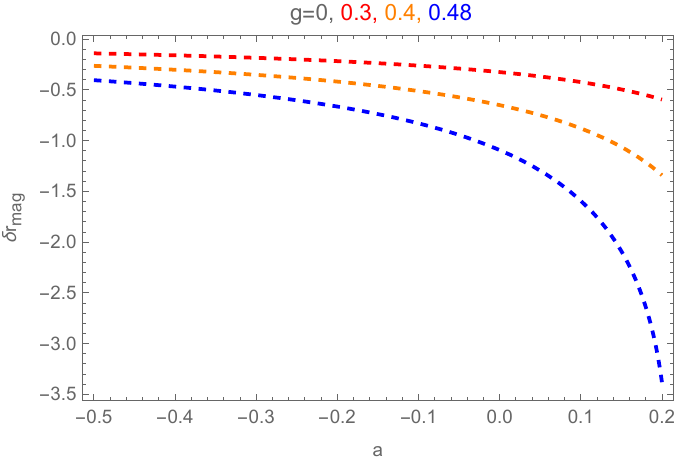}\\
         \includegraphics[width=0.25\linewidth]{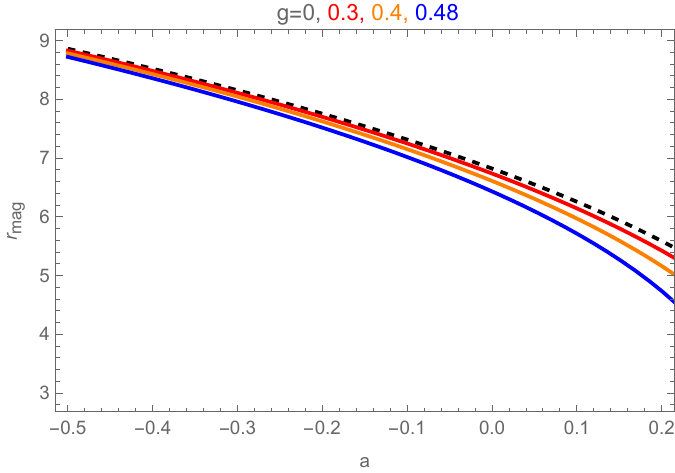}\hspace{1cm}
     \includegraphics[width=0.26\linewidth]{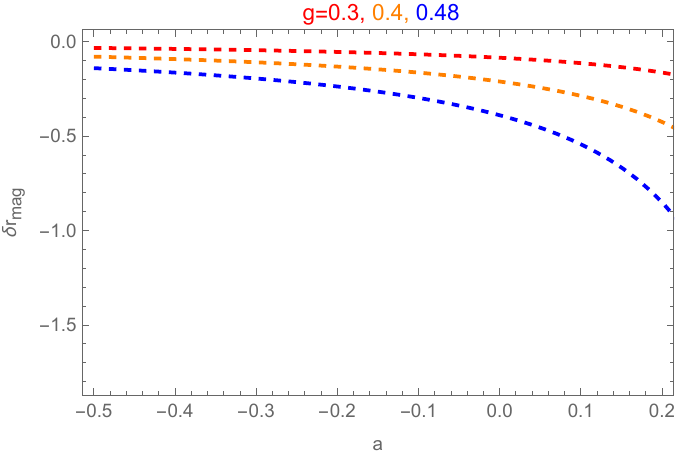}
    \caption{The behaviors of lensing observable $r_{mag}$ in strong gravitational lensing for the regular black hole with $\gamma= 2/3, n = 2$ (upper panel) and $\gamma = 1, n = 3 $ (lower panel).}
    \label{figa16}
\end{figure}

\begin{figure}[ht]
    \centering
    \includegraphics[width=0.25\linewidth]{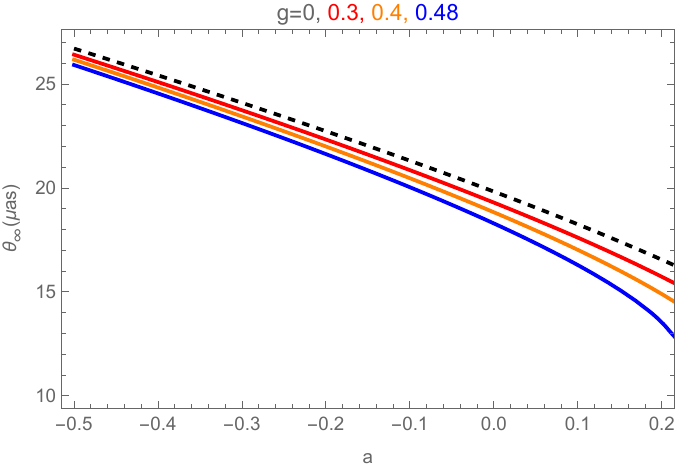}\hspace{1cm}
     \includegraphics[width=0.25\linewidth]{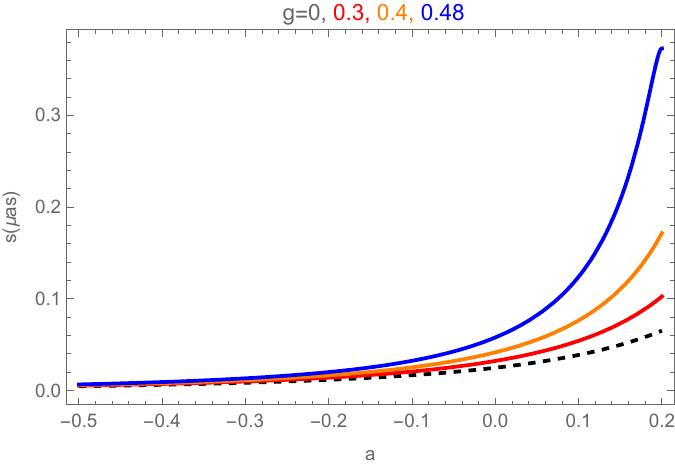}\\
          \includegraphics[width=0.25\linewidth]{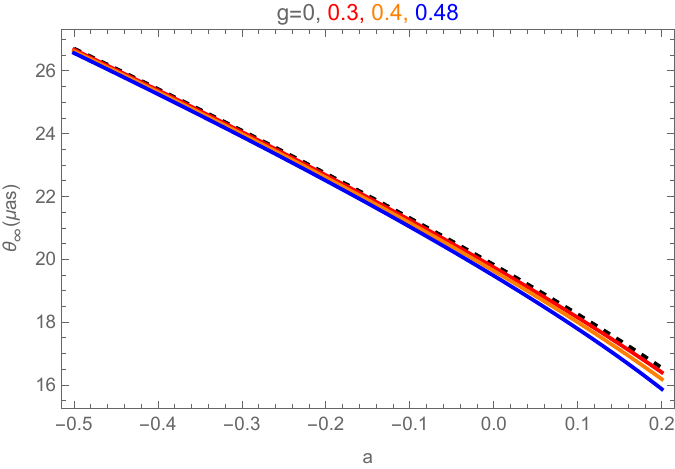}\hspace{1cm}
     \includegraphics[width=0.25\linewidth]{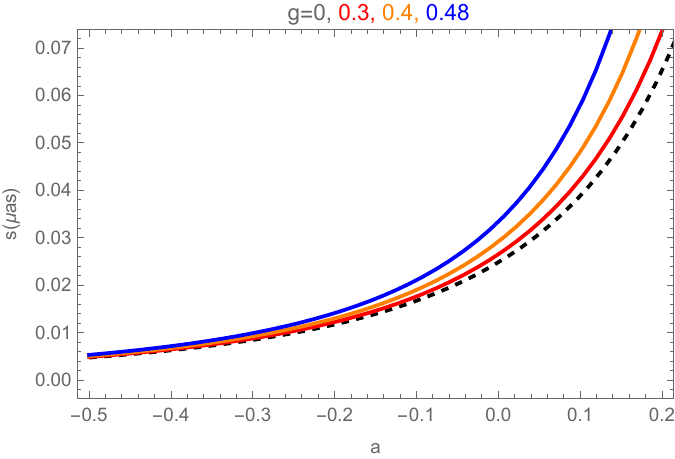}
    \caption{The behaviors of lensing observables $\theta_{\infty}$ and $s$ in strong gravitational lensing by considering the regular black hole with $\gamma= 2/3, n = 2$ (upper panel) and $\gamma = 1, n = 3 $ (lower panel) as the M87* black hole.}
    \label{figa17}
\end{figure}

\begin{figure}[ht]
    \centering
    \includegraphics[width=0.25\linewidth]{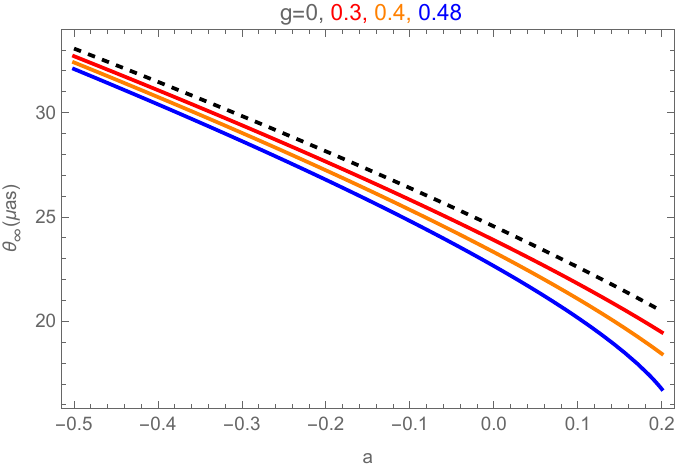}\hspace{1cm}
     \includegraphics[width=0.25\linewidth]{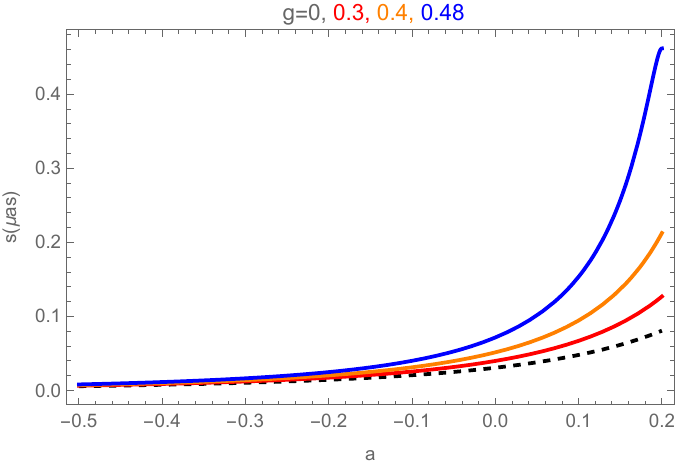}\\
          \includegraphics[width=0.25\linewidth]{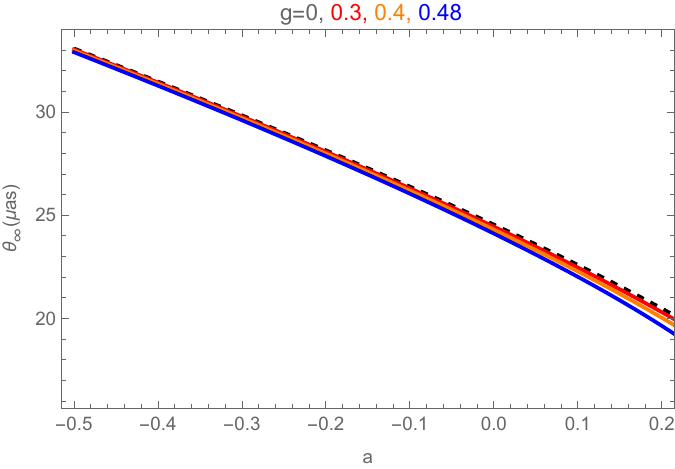}\hspace{1cm}
     \includegraphics[width=0.25\linewidth]{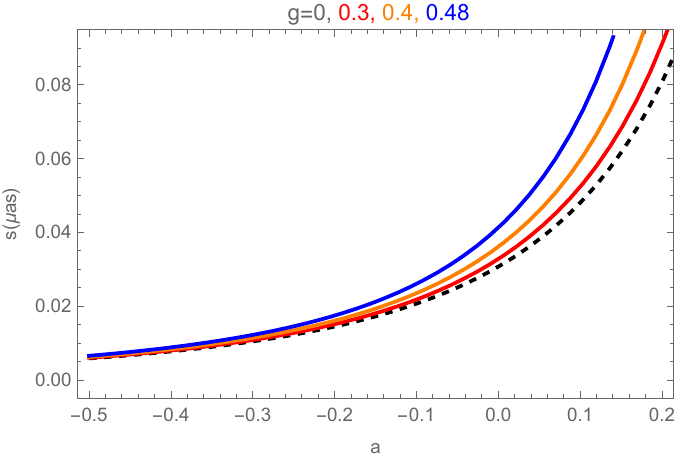}
    \caption{The behaviors of lensing observables $\theta_{\infty}$ and $s$ in strong gravitational lensing by considering the regular black hole with $\gamma= 2/3, n = 2$ (upper panel) and $\gamma = 1, n = 3 $ (lower panel) as the SgrA* black hole.}
    \label{figa18}
\end{figure}

Furthermore, we can also consider another crucial observable in strong-field lensing, the time delay defined as the difference in time between the formation of relativistic images. Since the time elapsed along the optical path for different images varies, so there exists a time delay effect among them. When the $p$-th and $q$-th images are located on the same side of the lens, their time delay can be approximated as \cite{Bozza:2003cp}
\begin{equation}
\Delta T_{p, q} \approx 2 \pi(p-q) \frac{\widetilde{R}\left(0, x_m\right)}{2 \bar{a} \sqrt{\mathfrak{n}_m}}+2 \sqrt{\frac{A_m u_m}{B_m}}[e^{(\bar{b}-2 q \pi \pm \beta) / 2 \bar{a}}-e^{(\bar{b}-2 p \pi \pm \beta) / 2 \bar{a}}],\label{eq:71}
\end{equation}
where
\begin{equation}
\widetilde{R}\left(z, x_m\right)=\frac{2\left(1-A_0\right) \sqrt{B A_0}[2 C-u D]}{A^{\prime} \sqrt{C\left(D^2+4 A C\right)}}\left(1-\frac{1}{\sqrt{A_0} f\left(z, x_0\right)}\right).\label{eq:72}
\end{equation}
As addressed in \cite{Bozza:2003cp}, the result mainly depends  on the first term, so this time delay can be roughly evaluated by
\begin{equation}
\Delta T_{p, q} \approx 2 \pi(p-q) \frac{\widetilde{R}\left(0, x_m\right)}{2 \bar{a} \sqrt{\mathfrak{n}_m}}=2 \pi(p-q) \frac{\tilde{a}}{\bar{a}}, \label{eq:73}
\end{equation}
where
\begin{equation}
\tilde{a}=\frac{\tilde{R}(0,x_m)}{2\sqrt{n_m}}. \label{eq:74}
\end{equation}
In addition, the propagation time of a prograde photon $(a>0)$ differs from that of a retrograde photon $(a<0)$, and we need to consider the case when the two images are located on opposite sides of the lens. In this case, the time delays are given by:
\begin{equation}
\Delta \widetilde{T}_{p, q} \approx \frac{\tilde{a}(a)}{\bar{a}(a)}[2 \pi p+\beta-\bar{b}(a)]+\tilde{b}(a)-\frac{\tilde{a}(-a)}{\bar{a}(-a)}[2 \pi q+\beta-\bar{b}(-a)]-\tilde{b}(-a),  \label{eq:75}
\end{equation}
where
\begin{equation}
\tilde{b}=-\pi+\tilde{b}_D\left(x_m\right)+\tilde{b}_R\left(x_m\right)+\tilde{a} \log \left(\frac{c x_m^2}{u_m}\right),  \label{eq:76}
\end{equation}
with
\begin{equation}
\tilde{b}_D=2 \tilde{a} \log \frac{2\left(1-A_m\right)}{A_m^{\prime} x_m}, \quad \tilde{b}_R\left(x_m\right)=\int_0^1\left[\widetilde{R}\left(z, x_m\right) f\left(z, x_m\right)-\widetilde{R}\left(0, x_m\right) f_0\left(z, x_m\right)\right] d z . \label{eq:77}
\end{equation}
{Thus, by recalling the lensing coefficients $\bar a$ and $\bar b$, and further calculating the coefficients $\tilde a$ and $\tilde b$, we can get the time delay between arbitrary two images.}

\subsection{ Evaluating the observables by supermassive black holes}
In this subsection, we shall examine the rotating regular black hole by considering it as the supermassive M87* and SgrA* black holes. We conduct numerical investigations on the lensing observables, namely $r_{\text{mag}}$, $\theta_{\infty}$, $s$ and time delays. For M87* black hole, we take the mass $M = 6.5 \times 10^9 M_\odot$ and the luminosity distance $D_{OL} = 16.8M_{ pc}$\cite{EventHorizonTelescope:2019ths} , and for SgrA* black hole, we take $M = 4.0 \times 10^6 M_\odot$ and $ D_{OL} = 8.35K_{pc}$ \cite{Chen:2019tdb}.

Fig.\ref{figa16} illustrates the dependence of $r_{mag }$ on the spin $a$ of black hole for various values of $g$ in the context of regular black holes with $\gamma= 2/3, n = 2$ and $\gamma = 1, n = 3 $. The plot shows that $r_{mag}$ decreases monotonically with increasing spin. Notably, the presence of the parameter $g$ leads to a reduction in $r_{mag}$ compared to the Kerr black hole scenario. When comparing the two types of regular black holes, it becomes apparent that for the regular black hole with $\gamma= 2/3, n = 2$ , the variation of $r_{mag }$ with the parameter $g$ is much more pronounced, particularly when $a > 0$. This phenomenon is similar to what we observed in the deflection angle of light.

\begin{figure}[ht]
    \centering
    \includegraphics[width=0.25\linewidth]{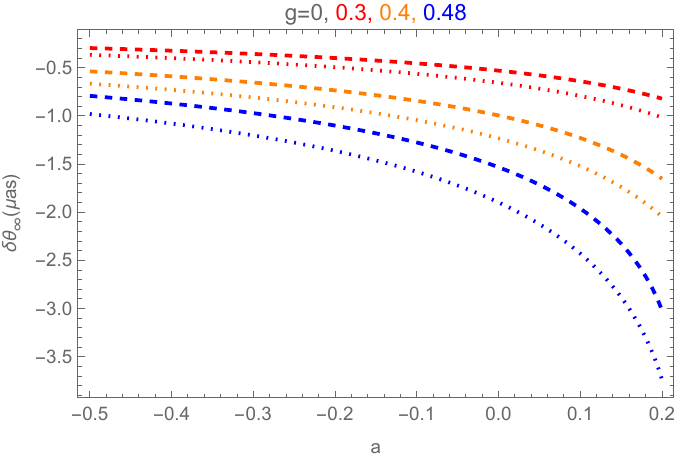}\hspace{1cm}
     \includegraphics[width=0.25\linewidth]{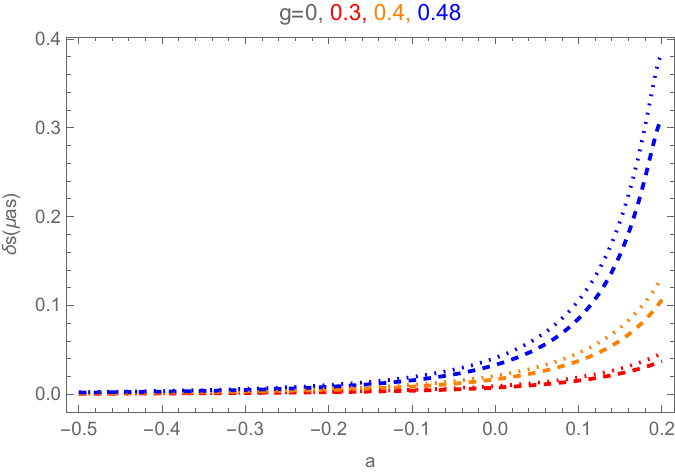}\\
          \includegraphics[width=0.25\linewidth]{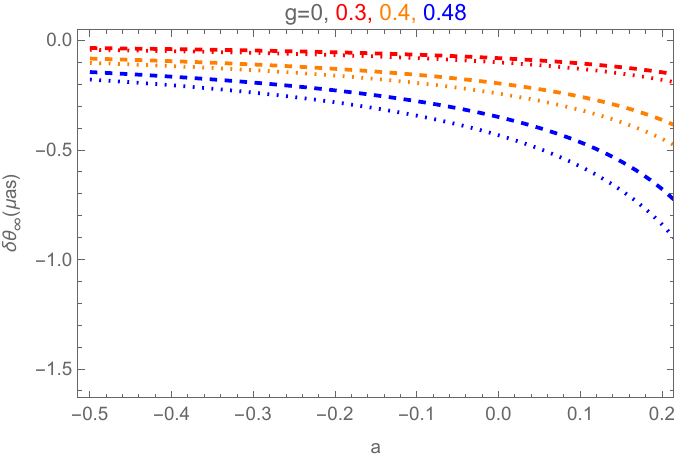}\hspace{1cm}
     \includegraphics[width=0.25\linewidth]{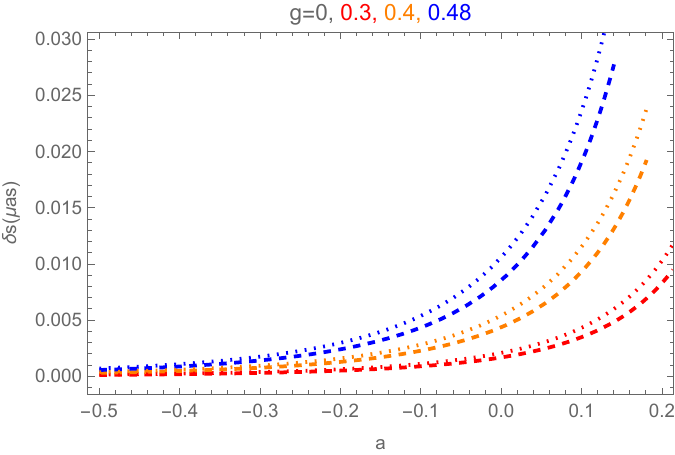}
    \caption{The differences of $\theta_{\infty}$ and $s$ between the regular black hole with $\gamma= 2/3, n = 2$ (upper panel) and $\gamma = 1, n = 3 $ (lower panel) and the Kerr black hole, where the dashed curves are for M87* while the dotted curves are for SgrA*.}
    \label{figa19}
\end{figure}

In the following analysis, we compute the values of $\theta_{\infty}$ and $s$ for both supermassive regular black holes as illustrated in Fig.\ref{figa17} (M87*) and Fig.\ref{figa18} (SgrA*), respectively. It is evident that the $\theta_{\infty}$ decreases for both M87* and SgrA* with increasing spin $a$, while the parameter $g$ contributes to the reduction in $\theta_{\infty}$. Conversely, $s$ shows an upward trend with increasing spin $a$, with the parameter $g$ enhancing its magnitude.
In Fig.\ref{figa19}, the deviations $\delta \theta_{\infty}$ and $\delta s$ from the Kerr black hole in both cases are depicted, where the dashed curves denote supermassive M87* black hole while the dotted curves denote the SgrA* black hole. We observe that for the regular black holes, both $\delta \theta_{\infty}$ and $\delta s$ are amplified by the parameter $g$. Moreover, as the spin parameter $a$ increases, both $\delta \theta_{\infty}$ and $\delta s$ also increase. Additionally, the deviations for M87* consistently exceed those for SgrA*. In the case of prograde photons, we find that the regular black hole with $\gamma=2/3$ and $n=2$ have larger variations of $\delta \theta_{\infty}$ and $\delta s$ than those with $\gamma=1$ and $n=3$.
Though the increase in $g$ increases the lensing observational deviations from Kerr black hole, which is around or even less than $1 \mu a s$, these deviations are much smaller than the resolution of the current EHT \cite{EventHorizonTelescope:2022xnr}, making it difficult to be observed. However, advancements in technology, including the new generation of Event Horizon Telescopes \cite{Ayzenberg:2023hfw}, hold the potential observational feasibility to distinguish the relativistic images and acquire lens coefficients.

\begin{table}[!ht]
{\centering
\begin{tabular}{|c|c|c|c|c|c|c|c|}
       \hline
\diagbox{a}{$\Delta T_{2,1}\left(\delta \Delta T_{2,1}\right) / \mathrm{hrs}$}{$g$} & 0 & 0.3 & 0.4 & 0.48\\
  \hline
       -0.2   &      332.220    &       326.379 (-5.841)        &        321.474(-10.746)     &   316.134 (-16.086)      \\
   \hline
       -0.1   &      311.480    &      304.833(-6.647)        &     299.151(-12.329)    &     292.831(-18.649)       \\
   \hline
       0      &     289.760    &        282.011(-7.749)      &      275.21(-14.550)     &    267.36(-22.400)       \\
    \hline
       0.1   &      266.734    &       257.376(-9.358)     &       248.776(-17.958)     &     238.076(-28.658)       \\
    \hline
       0.2   &      241.856    &     229.858(-11.998)       &       217.713(-24.143)    &     197.749(-44.107)       \\
    \hline
\end{tabular}
\caption{
\label{table1}The time delay $\Delta T_{2,1}$ between the first and second images for the rotating regular supermassive M87* black hole ($\gamma=2/3$ and $n=2$ ) and its deviation $\delta \Delta T_{2,1}$ from the 
Kerr black hole. }}
\end{table}

\begin{table}[!ht]
{\centering
\begin{tabular}{|c|c|c|c|c|c|c|c|}
       \hline
\diagbox{a}{$\Delta T_{2,1}\left(\delta \Delta T_{2,1}\right) / \mathrm{hrs}$}{$g$} & 0 & 0.3 & 0.4 & 0.48\\
  \hline
       -0.2   &      332.220    &       331.433(-0.787)       &        330.329(-1.891)     &    328.891 (-3.329)      \\
   \hline
       -0.1   &      311.480    &        310.532(-0.948)       &      309.192 (-2.288)    &     307.431(-4.049)       \\
   \hline
       0      &     289.760    &        288.582(-1.178)      &       286.9(-2.860)     &    284.657(-5.103)       \\
    \hline
       0.1   &      266.734    &       265.206(-1.528)     &        262.989(-3.745)     &     259.951(-6.783)       \\
    \hline
       0.2   &      241.856    &      239.736(-2.120)       &       236.561(-5.295)    &     231.952(-9.904)       \\
    \hline
\end{tabular}
\caption{
\label{table2}The time delay $\Delta T_{2,1}$ between the first and second images for the rotating regular supermassive M87* black hole ($\gamma=1$ and $n=3$ ) and its deviation $\delta \Delta T_{2,1}$ from the Kerr case.}}
\end{table}

\begin{table}[!ht]
{\centering
\begin{tabular}{|c|c|c|c|c|c|c|c|}
       \hline
\diagbox{a}{$\Delta T_{2,1}\left(\delta \Delta T_{2,1}\right) / \mathrm{min}$}{$g$}& 0 & 0.3 & 0.4 & 0.48\\
  \hline
       -0.2   &      12.267    &       12.051 (-0.216)      &     11.870(-0.397)     &   11.673 (-0.594)      \\
   \hline
       -0.1   &      11.501    &        11.256(-0.245)       &      11.046(-0.455)    &    10.812(-0.689)       \\
   \hline
       0      &     10.699    &       10.413(-0.286)      &    10.162(-0.537)     &     9.872(-0.827)       \\
    \hline
       0.1   &      9.849    &    9.503(-0.346)    &      9.186(-0.663)     &     8.791(-1.058)       \\
    \hline
       0.2   &      8.930    &      8.487(-0.443)        &        8.039(-0.891)    &   7.301(-1.629)       \\
    \hline
\end{tabular}
\caption{
\label{table3}The time delay $\Delta T_{2,1}$ between the first and second images for the rotating regular supermassive SgrA* black hole ($\gamma=2/3$ and $n=2$ ) and its deviation $\delta \Delta T_{2,1}$ from the 
Kerr black hole.}}
\end{table}

\begin{table}[!ht]
{\centering
\begin{tabular}{|c|c|c|c|c|c|c|c|}
       \hline
\diagbox{a}{$\Delta T_{2,1}\left(\delta \Delta T_{2,1}\right) / \mathrm{min}$}{$g$} & 0 & 0.3 & 0.4 & 0.48\\
  \hline
       -0.2   &      12.267    &       12.238(-0.029)       &      12.197(-0.070)     &   12.144 (-0.123)      \\
   \hline
       -0.1   &      11.501    &        11.466(-0.035)       &      11.416 (-0.845)    &     11.351(-0.150)       \\
   \hline
       0      &     10.699    &        10.655(-0.044)      &      10.593(-0.106)     &     10.510(-0.189)       \\
    \hline
       0.1   &      9.849    &      9.792(-0.057)     &       9.710(-0.139)     &     9.598(-0.251)       \\
    \hline
       0.2   &      8.930    &      8.852(-0.078)       &        8.735(-0.195)    &    8.564(-0.366)       \\
    \hline
\end{tabular}
\caption{
\label{table4}The time delay $\Delta T_{2,1}$ between the first and second images for the rotating regular supermassive SgrA* black hole ($\gamma=1$ and $n=3$ ) and its deviation $\delta \Delta T_{2,1}$ from the Kerr case.}}
\end{table}

The time delay $\Delta T_{2,1}$ between the first image and the second image of the regular black hole with $\gamma = 2/3, n = 2$ (Tables \ref{table1} for M87* and Tables \ref{table3} for SgrA*) and $\gamma = 1, n = 3$ (Tables \ref{table2} for M87* and Tables \ref{table4} for SgrA*), along with its deviation $\delta \Delta T_{2,1}$ from the Kerr black hole, are presented.
For each case, we observe that the values of $\Delta T_{2,1}$ for regular black holes are always shorter than those for the Kerr black hole, such that $\delta \Delta T_{2,1}$ is negative and its absolute value is enhanced by $g$ in all cases. Furthermore, the impact of spin on $\Delta T_{2,1}\left(\delta \Delta T_{2,1}\right)$ for the regular black holes mirrors that occurs in the Kerr case, with $\Delta T_{2,1}\left(\delta \Delta T_{2,1}\right)$ decreasing as spin increases. Notably, by comparing Tables \ref{table1}-\ref{table2} to Tables \ref{table3}-\ref{table4}, we see that the $\Delta T_{2,1}\left(\delta \Delta T_{2,1}\right)$ of M87* is significantly larger than that of SgrA*, which is reasonable because M87* is much far away. In addition, for both supermassive M87* and SgrA* black holes, we observe a pronounced increase in the changes of time delay $\Delta T_{2,1}$ and its deviation $\delta \Delta T_{2,1}$ for the case with $\gamma = 2/3, n = 2$, as compared to the configuration with $\gamma = 1, n = 3$.

In Tables \ref{table5}-\ref{table8}, we compute the time delay $\Delta \tilde T_{1,1}$ between prograde and retrograde images along with its deviation $\left(\delta \Delta T_{1,1}\right)$
for both regular black holes as M87* black hole (Tables \ref{table5}-\ref{table6}) and SgrA* black hole (Tables \ref{table7}-\ref{table8}).
As evident from the tables, when $a=0$, the time delay vanishes because of the spherical symmetry. However, its magnitude increases with faster rotation, and there exists symmetry between prograde and retrograde photons. Additionally, with an increase in $g$, $\Delta \tilde T_{1,1}\left(\delta \Delta \tilde T_{1,1}\right)$ in all cases exhibits a consistent growth trend. Furthermore, upon comparing the values in the tables, it becomes apparent that the time delay for M87* and its deviation can extend to hundreds (or tens) of hours, significantly exceeding the few minutes observed for SgrA*. 
Moreover, our results show that the time delay $\Delta \tilde T_{1,1}$ and its deviation $\delta \Delta T_{1,1}$ exhibit more change for the regular black hole with $\gamma = 2/3, n = 2$ as the parameter $g$ increases, contrasting the behavior of those with $\gamma=1, n=3$.

\begin{table}[!ht]
{\centering
\begin{tabular}{|c|c|c|c|c|c|c|c|}
       \hline
\diagbox{a}{$\Delta \tilde T_{1,1}\left(\delta \Delta \tilde T_{1,1}\right) / \mathrm{hrs}$}{$g$} & 0 & 0.3 & 0.4 & 0.48\\
  \hline
       -0.2   &      111.564    &     120.506(8.942)          &     131.921(20.357)     &   162.83(51.266)      \\
   \hline
       -0.1   &     55.355    &       59.308(3.953)       &     63.838 (8.483)    &    71.388(16.033)       \\
   \hline
       0      &          0    &                   0          &              0           &     0       \\
    \hline
       0.1   &       -55.355    &     -59.308(-3.953)         &    -63.838 (-8.483)    &    -71.388(-16.033)        \\
    \hline
       0.2   &         -111.564    &     -120.506(-8.942)          &    -131.921(-20.357)     &   -162.83(-51.266)      \\
    \hline
\end{tabular}
\caption{
\label{table5} The time delay $\Delta \tilde T_{1,1}$ between the prograde and retrograde images of the same order for rotating regular supermassive M87* black hole ($\gamma=2/3$ and $n=2$ ) and its deviation $\delta \Delta \tilde T_{1,1}$ from the Kerr case.}}
\end{table}

\begin{table}[!ht]
{\centering
\begin{tabular}{|c|c|c|c|c|c|c|c|}
       \hline
\diagbox{a}{$\Delta\tilde T_{1,1}\left(\delta \Delta\tilde T_{1,1}\right) / \mathrm{hrs}$}{$g$} & 0 & 0.3 & 0.4 & 0.48\\
  \hline
       -0.2   &      111.564    &      113.652(2.088)       &      116.995(5.431)     &     122.401(10.837)      \\
   \hline
       -0.1   &     55.355    &       56.269(0.914)       &     57.686(2.331)    &     59.831(4.476)       \\
   \hline
       0      &          0    &                   0          &              0           &     0       \\
    \hline
       0.1   &       -55.355    &        -56.269(-0.914)       &     -57.686(-2.331)    &     -59.831(-4.476)     \\
    \hline
       0.2   &         -111.564    &    -113.652(-2.088)       &      -116.995(-5.431)     &     -122.401(-10.837)     \\
    \hline
\end{tabular}
\caption{
\label{table6}The time delay $\Delta \tilde T_{1,1}$ between the prograde and retrograde images of the same order for rotating regular supermassive M87* black hole($\gamma=1$ and $n=3$ ) and its deviation $\delta \Delta\tilde T_{1,1}$ from the Kerr case.}}
\end{table}

\begin{table}[!ht]
{\centering
\begin{tabular}{|c|c|c|c|c|c|c|c|}
       \hline
\diagbox{a}{$\Delta \tilde T_{1,1}\left(\delta \Delta \tilde T_{1,1}\right) / \mathrm{min}$}{$g$} & 0 & 0.3 & 0.4 & 0.48\\
  \hline
       -0.2   &      4.119    &     4.449(0.330)       &      4.871(0.752)     &    6.012(1.893)      \\
   \hline
       -0.1   &     2.044    &       2.190(0.146)       &     2.357 (0.313)    &     2.636(0.592)       \\
   \hline
       0      &          0    &                   0          &              0           &     0       \\
    \hline
       0.1    &     -2.044    &       -2.190(-0.146)       &     -2.357 (-0.313)    &     -2.636(-0.592)         \\
    \hline
       0.2   &      -4.119    &      -4.449(-0.330)       &      -4.871(-0.752)     &    -6.012(-1.893)     \\
    \hline
\end{tabular}
\caption{ The time delay $\Delta \tilde T_{1,1}$ between the prograde and retrograde images of the same order for rotating regular supermassive SgrA* black hole ($\gamma=2/3$ and $n=2$ ) and its deviation $\delta \Delta \tilde T_{1,1}$ from the Kerr black hole.
\label{table7} }}
\end{table}

\begin{table}[!ht]
{\centering
\begin{tabular}{|c|c|c|c|c|c|c|c|}
       \hline
\diagbox{a}{$\Delta\tilde T_{1,1}\left(\delta \Delta\tilde T_{1,1}\right) / \mathrm{min}$}{$g$} & 0 & 0.3 & 0.4 & 0.48\\
  \hline
       -0.2   &      4.119    &     4.196(0.077)       &      4.420(0.201)     &    4.519(0.400)      \\
   \hline
       -0.1   &     2.044    &       2.078(0.034)       &     2.130 (0.086)    &     2.209(0.165)       \\
   \hline
       0      &          0    &                   0          &              0           &     0       \\
    \hline
       0.1    &     -2.044    &      -2.078(-0.034)       &     -2.130 (-0.086)    &     -2.210(-0.165)         \\
    \hline
       0.2   &      -4.119    &      -4.196(-0.077)       &      -4.320(-0.201)     &    -4.519(-0.400)     \\
    \hline
\end{tabular}
\caption{ The time delay $\Delta \widetilde{T}_{1,1}$ between the prograde and retrograde images of the same order for the rotating regular supermassive SgrA* black hole ($\gamma=1$ and $n=3$ ) and its deviation $\delta \Delta \widetilde{T}_{1,1}$ from the Kerr case.
\label{table8} }}
\end{table}

\section{Conclusion and discussion}

In this paper, we concentrated on two classes of rotating regular black holes with Minkowski core ($\gamma = 2/3, n=2$ and $\gamma = 1, n=3$) and examined the strong gravitational lensing effects. Before studying the lensing observations, we started from the null geodesic and calculated the light deflection angle in strong field regime. Our results show that the deflection angle for the rotating regular black holes is always smaller than that for Kerr black hole. Then we combined the deflection angle and lens equation to formulate various lensing observations.

Firstly, we studied the lensing observables in space, namely the angular position of the asymptotic image ($\theta_{\infty}$), the angular distance between the outermost image and the asymptotic image ($s$) and the relative magnification of the outermost image compared to other relativistic images ($r_{\text{mag}}$). We mainly investigated the effect of the regular black hole parameter $g$ on the lensing observables and analyzed those deviations from Kerr black hole. Our analysis indicated that the parameter $g$ leads to a decrease in $\theta_{\infty}$ and $r_{\text{mag}}$, while increasing $s$. In the case of retrograde photons, the effects are more pronounced for the regular black holes with $\gamma = 2/3, n = 2$. 
Furthermore, the deviations of the observables between the regular black hole with Minkowski core and the Kerr black hole are more pronounced in the case of M87* than in SgrA*.
Although these deviations are difficult to be detected by the current EHT, we can expect that the resolution in next-generation Event Horizon Telescope are precise enough to observe the relativistic images and distinguish the deviations from Kerr black hole.

Next, we studied the time delay of relativistic image, namely, $\Delta T_{2,1}$ for the first image and the second image, as well as $\Delta \tilde T_{1,1}$ for the prograde and retrograde images. We found that the $\Delta \tilde{T}_{2,1}$ values for regular black holes with Minkowski core are always shorter than those for Kerr black holes. Conversely, the $\Delta \tilde{T}_{1,1}$ values for regular black holes are longer than those for Kerr black holes. 
In addition, the time delay and its deviation for M87* are notably longer in comparison to SgrA*, suggesting that M87* could be a more favorable gravitational center for detecting signatures of regular black holes through time delay measurements.
The time delay displays a more pronounced influence on regular black holes with $\gamma = 2/3, n = 2$, contrasting their behavior to those with $\gamma = 1, n = 3$. Consequently, the regular black hole with $\gamma = 2/3, n = 2$ is relatively easier to identify in the point view of time delay. 
Although the deviations of lensing observables from Kerr black hole are challenging to be detected with the current instruments, our findings indicate that there are notable and detectable difference in the time delays from Kerr black hole. Therefore, if future observations can distinguish between different images, the time delay may serve as a potential indicator for differentiating regular black holes from Kerr black holes.

In conclusion, when considering rotating regular black hole with Minkowski cores as potential candidates for supermassive black holes, namely in M87* and SgrA*, our results indicate that strong gravitational lensing can distinguish these black holes from Kerr black holes by discerning differences in theoretical predictions, physical processes, and formation scenarios.Specifically, the parameter $g$ shifts the maximum Kretschmann scalar to a larger radius and affects the Hawking temperature, indicating changes in thermodynamic properties and black hole stability. It is interesting to study other observational phenomena, such as the nearby orbits and gravitational waves, of which the outcomes might complement our lensing studies to further help us understand the theoretical and observable properties of regular black holes with Minkowski cores. Additionally, since regular black holes have different classification as we mentioned in the introduction, another future direction is to extend the current studies and related topics to other classes of regular black holes, which are important to extract potential universal features of regular black holes, especially those different from normal black holes with singularity.

\section*{Acknowledgments}
We are very grateful to Prof. Yi Ling and Prof. Xin Wu for helpful discussions. 
This work is partly supported by the Natural Science Foundation of China under Grants No. 12375054 and 12405067.
It is also supported by the financial support from Brazilian agencies Funda\c{c}\~ao de Amparo \`a Pesquisa do Estado de S\~ao Paulo (FAPESP), Funda\c{c}\~ao de Amparo \`a Pesquisa do Estado do Rio de Janeiro (FAPERJ), Conselho Nacional de Desenvolvimento Cient\'{\i}fico e Tecnol\'ogico (CNPq), and Coordena\c{c}\~ao de Aperfei\c{c}oamento de Pessoal de N\'ivel Superior (CAPES).

\end{document}